\documentclass[10pt]{iopart}
\ioptwocol

\usepackage{iopams}  
\usepackage{graphicx}
\usepackage{graphics}
\usepackage{epsf}

\usepackage{amssymb}

\def\raE     {\rightarrow}

\def \Kp3pi  {${\rm K^+} \rightarrow \pi^+\pi^-\pi^+$~}
\def \Kp3pig {${\rm K^+} \rightarrow \pi^+\pi^-\pi^+(\gamma)$~}
\def \Kmunu  {${\rm K^{\pm}} \rightarrow \mu^{\pm}\nu(\gamma)$~}
\def \Kpipi  {${\rm K^{\pm}} \rightarrow \pi^{\pm}\pi^0(\gamma)$~}

\def\ifm#1{\relax\ifmmode#1\else$#1$\fi}

\def\fig#1{fig.~\ref{#1}}

\def\equ#1{eq.~\ref{#1}}
\def\Equ#1{Eq.~\ref{#1}}
\def\etal{{\it et al.}}

\def\f{\ifm{\Phi}}    

\def\epm{\ifm{e^+e^-}}
\renewcommand{\to}{\ensuremath{\rightarrow}}
\def\dafne{DA$\Phi$NE}

\def\ks{\ifm{K_S}}
\def\kl{\ifm{K_L}}

\def\kp{\ifm{K^+}}

\def\pip{\ifm{\pi^+}}
\def\pim{\ifm{\pi^-}}
\def\pio{\ifm{\pi^0}}
\def\JPC{\ifm{J^{PC}}}

\def\ab{\ifm{\sim}}
\newcommand{\ket}[1]{\ifm{|#1\rangle}}
\newcommand{\braket}[2]{\ifm{\langle #1|#2 \rangle}}
\def\order#1,{\ifm{\mathcal{O}(10^{#1})}}

\def\MeV{\ifm{\mbox{MeV}}}
\def\GeV{\ifm{\mbox{GeV}}}
\def\mum{\ifm{\mbox{$\mu$m}}}
\def\mm{\ifm{\mbox{mm}}}

\def\invfb{\ifm{\mbox{fb}^{-1}}}
\def\invpb{\ifm{\mbox{pb}^{-1}}}
\def\cms{\ifm{\mbox{cm}^{-2}}\mbox{s}^{-1}}

\def\Dt{\ifm{\Delta \tau}}

\def\Damu{\ifm{\Delta a_\mu}}
\def\Dax{\ifm{\Delta a_X}}
\def\Day{\ifm{\Delta a_Y}}
\def\Daz{\ifm{\Delta a_Z}}
\def\Dao{\ifm{\Delta a_0}}

\begin{document}

\title[]{DA$\Phi$NE \& KLOE-2}

\author{Antonio De Santis\footnote{on behalf of KLOE-2 Collaboration and \dafne\ team:
D. Alesini, M.E. Biagini, S. Bini, M. Boscolo, B. Buonomo, S. Cantarella,
G. Di Pirro, G. Delle Monache, A. Drago, L. Foggetta, O. Frasciello, A. Gallo, 
A. Ghigo, S. Guiducci, C. Ligi, G. Mazzitelli, C. Milardi, L. Pellegrino, R. Ricci, 
U. Rotundo, C. Sanelli, M. Serio, A. Stella, A. Stecchi, M. Zobov, LNF-INFN, Frascati, Italy.}}

\address{INFN - Laboratori Nazionali di Frascati, v. Enrico Fermi, 50, 00044}
\ead{antonio.desantis@lnf.infn.it}
\vspace{10pt}
\begin{indented}
\item[]February 2015
\end{indented}

\begin{abstract}
The DA$\Phi$NE collider, located in the Frascati National Laboratories of 
INFN, has two main rings, where electrons and positrons
are stored to collide at a center of mass energy of 1.02 GeV, the 
$\phi$ resonance mass.
KLOE-2 experiment is located at the collider interaction region. 
The detector is capable to observe and collect data coming from 
$\phi$ decay: charged and neutral kaon pairs, lighter unflavored mesons
($\eta$, $\eta'$, $f_0$, $a_o$, $\omega/\rho$).
%
%
\par
In the first half of 2013 the KLOE detector has been upgraded 
inserting new detector layers in the inner part of the apparatus, around 
the interaction region in order to improve detector hermeticity and acceptance. 
The long shutdown has been used to undertake a general consolidation program 
aimed at improving the \dafne\ performances.
%
%
\par
This contribution presents the $\phi$-factory setup and the achieved 
performances in terms of beam currents, luminosity and 
related aspects together with the KLOE-2 physics program, upgrade 
status report and recent physics results. 
\end{abstract}

\section{\dafne\ preparation for KLOE-2 physics run}
\dafne\ \cite{dafne1} is a double ring lepton collider working at the \f-resonance c.m. energy (1.02 GeV). 
The two rings layout crosses two times, but the vacuum chamber is shaped in a such way that only one 
Interaction Region (IR) is possible where the KLOE-2 detector is currently installed. 
The complex includes a S-band linac, 180 m long transfer lines and an accumulator/damping ring, 
providing fast and high efficiency top-up electron/positron injection. 
\dafne\ reached its maximum luminosity during the test of the new Crab-Waist (CW) collision scheme 
\cite{dafne2} achieving a peak luminosity, $L = 4.5\times 10^{32} \cms$ , two orders of magnitude
higher than the one measured by other colliders working at the same c.m. energy.

The new IR \cite{dafne3} for the KLOE-2 experiment with CW has been designed, installed and 
tested already in the 2010-2012 reaching a instantaneous luminosity of $L = 1.52\times 10^{32} \cms$ \cite{dafne4}.
During the first seven months of 2013, the accelerator complex has been shut-down again mainly to 
install KLOE-2 detector upgrade \cite{dafne5} and to implement a general consolidation program 
concerning several machine subsystems \cite{dafne6,dafne7}. Such upgrade imposed the extraction of 
the IR from inside the detector and the disassembly of the low-$\beta$ section thus implying new 
commissioning \cite{dafne8}.  

\begin{figure}[htb]
   \begin{center}
     \includegraphics[width=0.40\textwidth]{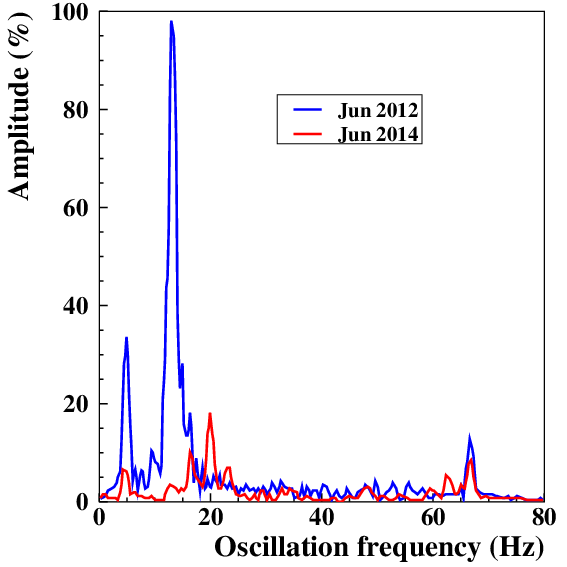}
   \end{center}
   \caption{IR mechanical vibration spectrum. 
     Darker line is for the new supporting structure.}
   \label{dafne:ir_vibration}
 \end{figure}

During the test run of 2012, the mechanical structure of the IR had shown to 
be inadequate to steadily support the heavy defocusing quadrupoles cantilevered 
inside the detector. As a consequence, the two
beams were oscillating in phase at 10 Hz in the vertical plane, as shown in \fig{dafne:ir_vibration}. 
Moreover some components 
got seriously damaged: some bellows in the IR, 
which had lost electrical continuity causing anomalous beam induced heating of
one of the two defocusing quadrupoles, resulting in a harmful random vertical tune-shift.

The vacuum chamber around the Interaction Point (IP) has been replaced. The new one has tapered transition
between the thin ALBEMET sphere and the Al beam pipes, and includes reshaped bellows with new designed
RF contacts. Replacing the bellows solved the low-$\beta$ defocusing quadrupole heating problems,
recovering working point stability in operations.

Two cooling pipes have been added on the tapers and new semi-cylindrical thin (35 \mum) beryllium shields have
been placed inside the sphere. Two additional Beam Position Monitors (BPM) have been installed on both
sides of the IP, for a more accurate beams overlap and to perform transverse betatron coupling studies.

During the \dafne\ long standing history the high positron current regime was always characterized 
by instabilities induced by the electron cloud. During the test run of 2012
special electrodes \cite{dafne11} inside dipoles and wigglers vacuum chambers were installed to mitigate 
the effect. Several measurements and tests demonstrated their effectiveness in thwarting the e-cloud \cite{dafne12}. 
These first studies have all been done by polarizing the strip-line with a positive voltage in the range 0-250 V while 
simulations indicate that a factor two higher voltage is required to completely neutralize the e-cloud density due to a 
positron current of the order of 1 A. 
For this reason during the 2013 shutdown the electrode power supplies have been replaced with devices providing a 
maximum negative voltage of 500 V. The change of polarity was intended to limit the current delivered by the 
power supplies. 

\section{\dafne\ commissioning}
\dafne\ operation started by the end of January 2014, but it has been severely slowed down by
three main interruptions due to external circumstances that imposed two and a half months of inactivity.

The betatron tunes working points adopted by now are: $\nu_x^-$ = 5.098, $\nu_y^-$ = 5.164 and 
$\nu_x^+$ = 5.1023, $\nu_y^+$ = 5.139, which, according LIFETRAC \cite{dafne9} simulations should provide
good luminosity and lay in a rather large stable area still to be explored.

Transverse betatron coupling has been optimized by tuning the rotations of the low-$\beta$ focusing quadrupoles. 
Presently a very good coupling correction has been achieved for the positron beam, $\kappa \sim 0.4\%$, without
skew quadrupoles, while a further optimization is needed for the electron beam.
Tuning the skew quadrupoles a transverse betatron coupling of 0.2\%-0.3\% has been achieved in both rings.

Machine operation at high current strongly depends on vacuum conditions. Since the beam pipe has been opened
in several main ring sections, a quite long time has been spent to recover a reasonable dynamic vacuum level.
Highest currents stored, so far, in 98 contiguous bunches are 1.7 A and 1.15 A, for electron and positron beam, 
respectively. These currents are the highest ever achieved after installing the new IR for the KLOE-2 detector, 
based on the CW collision scheme. To reach such high current the three independent bunch-by-bunch feedback 
systems \cite{dafne10} installed on each \dafne\ ring are essential. The positron vertical feedback 
is now using a new ultra-low noise front-end module, designed in collaboration with the SuperKEKB feedback team.

Presently beam dynamics in the positron ring is clearly dominated by the electron-cloud induced instabilities. 
To cope with these effects several system are used: bunch-by-bunch transverse feedback systems, solenoids wound 
all around the straight sections and on-purpose designed electrodes \cite{dafne11} installed inside dipole and wiggler 
vacuum chambers. 
The new setup of electrodes, described in the previous section, has been tested storing a $\sim700$ mA current in 90 
bunches spaced by 2.7 ns, and measuring the horizontal and vertical tune spread along the batch with the electrodes 
on and off. Results show a clear reduction of the tune spread in both planes, but especially in the horizontal 
one \cite{dafne13}.

\begin{figure}[htb]
   \begin{center}
     \includegraphics[width=0.50\textwidth]{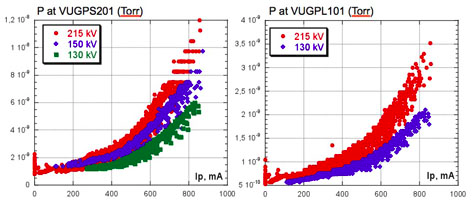}
   \end{center}
   \caption{Variation of the vacuum pressure in the arcs as a function of RF cavity voltage. The reduction
   of RF High Voltage results in a slower growth of the dynamical vacuum pressure because the induced 
   lengthening of the beam bunch reduces the charge density with same total beam current and thus the 
   e-cloud acceleration towards the beam line is weaker.}
   \label{dafne:rfhveffect}
 \end{figure}

Another positive result in mitigating the detrimental effects induced by the e-cloud has been obtained 
lengthening the bunch by reducing the voltage of the RF cavity of the positron ring, as shown in 
\fig{dafne:rfhveffect}, where the dynamic vacuum condition is shown as a function of the beam current 
for different High Voltage sets of the RF cavity. 

\begin{figure}[htb]
   \begin{center}
     \includegraphics[width=0.50\textwidth]{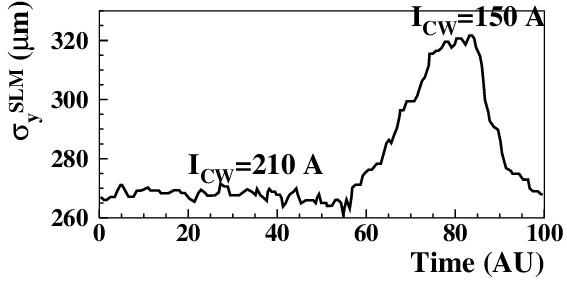}
   \end{center}
   \caption{Effect of the CW sextupoles on the transverse dimensions of the positron beam. When the CW
   sextupole current is lowered from 210 A to 150 A, a sizeable growth of vertical beam size is observed 
   at the Sincrotron Light Monitor from 265 \mum\ to 320 \mum.}
   \label{dafne:cweffect}
 \end{figure}

Collisions has been optimized without the CW Sextupoles. The overlap between colliding beams has
been recovered, both in transverse and longitudinal plan, taking advantage from the two new beam position
monitors installed around the IP. The highest luminosity measured in large Piwinsky angle regime without
the CW Sextupoles has been slightly in excess of $10^{32} \cms$. In this configuration several 
effects have been observed such as: considerable reduction of the lifetime of one beam when injecting the
other, transverse beam blow-up at high currents and deleterious background level. These phenomena are
correlated to the synchro-betatron resonances affecting collisions in large Piwinsky angle regime. 
The CW Sextupoles have been aligned on the collision orbit, by using beam based alignment
techniques, and their field increased progressively, while tuning collisions, till reaching strengths 
of the order of 70\% and 50\% of the nominal ones for positron and electron, respectively.
The reduction of the vertical beam size, as seen on the synchrotron light monitor, is shown 
in \fig{dafne:cweffect}.

The increase of bunch length, useful for mitigating the electron cloud effect on the positron beam,  
for both beams appreciably contributed in enhancing peak luminosity.
The convoluted vertical sizes of the colliding bunches as measured by beam-beam scan at low current is
$\sigma_y = 5.4 \mum$, due to the not yet optimized transverse coupling in the electron beam.
The highest luminosity attained, so far, is $L=1.71 \times 10^{32} \cms$ measured with current of the order of
$I^- \sim 0.96 \mbox{ A}$ and $I^+ \sim 0.89 \mbox{ A}$ stored in 100 bunches. 
This value exceeds by a 13\% the best luminosity ever achieved, at \dafne\, during operations  with KLOE detector.

Instantaneous luminosity in the present \dafne\ configuration is measured by the KLOE-2 detector every
30 seconds. A faster monitor measures the photon emitted at small angle ($\sim$1 mrad) in the $\epm$ inelastic scattering and is used only for relative luminosity measurements, during collision 
optimization. 

\begin{figure}[htb]
   \begin{center}
     \includegraphics[width=0.50\textwidth]{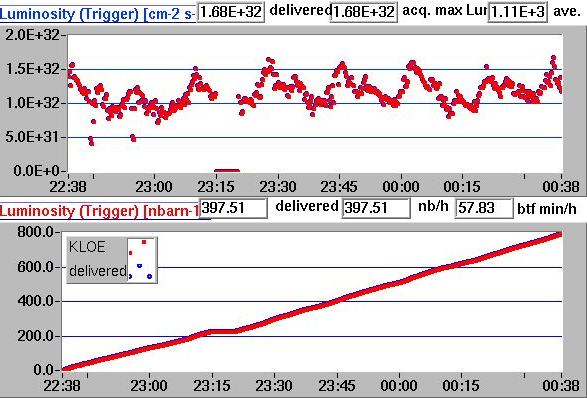}
   \end{center}
   \caption{Instantaneous (top) and integrated (bottom) luminosity.}
   \label{dafne:lumi}
 \end{figure}

Furthermore a rather promising hourly integrated luminosity has been recorded averaging over two hours,
in moderate injection regime: $\int_{1h}L \sim 0.4~\invpb$ , see \fig{dafne:lumi}
This measurement indicates that, having a reasonable up-time, a daily integrated luminosity at least of the order
of $\sim 10 \invpb$ can be delivered to the experiment.

\section{KLOE-2 experimental apparatus}

The KLOE-2 experiment is an upgraded version of the old KLOE apparatus
with the inclusion of new sub-detectors allowing for larger physics 
program with increased reconstruction performance.

The original KLOE detector consists of a large cylindrical drift chamber (DC) 
surrounded by a lead-scintillating fiber electromagnetic calorimeter (EMC). 
A super-conducting coil around the EMC provides a 0.52 T axial field. 
\par
The DC \cite{Adinolfi:2002uk} is 4 m in diameter and 3.3 m long and has 
12,582 all-stereo cells properly arranged in 58 layers to ensure homogeneous 
detector response. Time and amplitude of signals from cells are read-out 
to measure hit positions and energy loss. 
The chamber structure is made of carbon-fiber epoxy 
composite 
and the gas mixture used is 90\% helium, 10\% isobutane. These features 
maximize transparency to photons and reduce \kl\to\ks\ regeneration and multiple
scattering. The position resolutions for single hits are $\sigma_{r,\phi}$\ab 150
\mum\ and $\sigma_z$\ab 2 \mm\ in the transverse and longitudinal plane, 
respectively and are almost homogeneous in the active volume. The momentum 
resolution is $\sigma(p_{\perp})/p_{\perp}\ab 0.4\%$ for polar angles in the range 
$40^\circ-130^\circ$. 
\par
The EMC \cite{Adinolfi:2002jk} is divided into a barrel and two end-caps covering 
98\% of the solid angle. Signals from impinging particles are read out at both
ends of each module by photo-multipliers for a total of 2440 
cells arranged in five layers in depth. Amplitudes and time of the signals are recorded.
Cells close in time and space are grouped 
into calorimeter clusters. The cluster energy $E_\mathrm{clu}$ is the sum of its 
cell energies. 
The cluster time $T_\mathrm{clu}$ and position $\vec{R}_\mathrm{clu}$ 
are energy-weighted 
averages. Energy and time resolutions are $\sigma_E/E_\mathrm{clu} = 
5.7\%/\sqrt{E_\mathrm{clu}(\GeV)}$ and  
$\sigma_{T_\mathrm{clu}} = 57{\mathrm{ps}}/\sqrt{E_\mathrm{clu}(\GeV)} \oplus 100\ 
{\rm \mathrm{ps}}$, 
respectively. 
\par

The KLOE-2 experimental program \cite{AmelinoCamelia:2010me}  
implies several detector improvements in order to increase the 
physics results outcomes. 
For the gamma-gamma physics two pairs of electron-positron 
taggers have been installed. A small LYSO crystal calorimeter matrix, 
the Low Energy Tagger \cite{Babusci:2009sg} inside KLOE apparatus
and a plastic scintillator hodoscope, the High Energy Tagger
\cite{Archilli:2010zza}, along the beam lines outside the KLOE detector. 

To improve the acceptance and the angular coverage two new calorimeters 
have been developed.
A pair of LYSO crystal calorimeters (CCALT) \cite{Happacher:2009xm} have 
been installed near the interaction region to improve the angular acceptance 
for low-$\theta$ particles. This calorimeter will be also useful to provide
fast signals for luminosity measurement and beam instability feedback 
to help DAFNE tune-up. 

A pair of tile calorimeters (QCALT) \cite{Cordelli:2009xb}, covers the 
quadrupoles inside the KLOE detector and along the beam pipe. These 
calorimeters are made of tungsten slabs and singly read-out scintillator 
tiles to improve the angular coverage for particles coming from the active 
volume of the DC (e.g. \kl\ decay).

In order to increase the resolution on the vertex reconstruction for decay 
in the vicinity of the primary interaction point a small and light inner 
tracker (IT) \cite{Balla:2010jr} made of four planes of cylindrical GEM 
has been designed.
\par
During data taking \dafne\ beam conditions and detector calibrations are 
constantly monitored in order to guarantee the highest quality of the 
collected data.

\section{Physics result with KLOE}

\subsection{Kaon interferometry}\label{ads:sec:intro}

The \f\ meson produced at \dafne\ \f-factory have a small residual 
momentum in the horizontal plane and the branching fraction for the
\f\ decay in neutral kaon pair is 34\%. 
This decay, due to strong interaction, preserve parity (P) and charge
conjugation (C) eigenvalue of the initial state: $\JPC=1^{--}$.
The initial state can be represented as an anti-symmetric combination 
of the two neutral kaon mass eigenstate:
$$
\ket{\f} \propto \ket{\ks,\vec{p}}\ket{\kl,-\vec{p}}-\ket{\ks,-\vec{p}}\ket{\kl,\vec{p}}
$$ 
The time evolution of the kaon system preserve the initial 
correlation even at the kaon decay level. 
Labelling $f_1$ and $f_2$ the final decay channels 
for the two kaons and evaluating the probability of a decay into 
\ket{f_1,f_2} final state as a function of the difference of proper 
decay times ($\Dt = \tau_2-\tau_1$) the following equation is obtained 
(For a detailed treatment of quantum interference \cite{DiDomenico:2007zza}).
\begin{equation}\label{eq:intro:timeevo} 
  I_{f_1f_2}(\Dt)  \propto  e^{-\Gamma |\Dt|} \Big[ 
    |\eta_{f_1}|^2 e^{\frac{\Delta\Gamma}{2} \Dt} + 
    |\eta_{f_2}|^2 e^{-\frac{\Delta \Gamma}{2}\Dt}
\end{equation}
$$
-2 \Re e     \Big( \eta_{f_1} \eta_{f_2}^* e^{-i\Delta m \Dt} \Big) \Big]
$$
where $ \eta_{f_j}=\braket{f_j}{\kl}/\braket{f_j}{\ks}$, 
$\Gamma = \Gamma_S+\Gamma_L$ and $\Delta \Gamma = \Gamma_S -\Gamma_L$.
\par
\Equ{eq:intro:timeevo} shows a time interference term characteristic of 
the so-called EPR correlation from the names of Einstein, Podolsky and Rosen 
\cite{Einstein:1935rr} that firstly pointed out this kind of effect.
\par
In this analysis the final state chosen is 
$\ket{f_1}=\ket{f_2}=\ket{\pip\pim}$ and a fully destructive interference 
is expected for equal decay times ($|\Dt| = 0$). 
The ratio of neutral kaon decay amplitudes ($\eta_j$) becomes: 
\begin{equation} \label{eq:intro:etapm}
  \eta_j = \eta_{\pip\pim} = 
  \frac{\braket{\pip\pim}{T|\kl}}{\braket{\pip\pim}{T|\ks}} \simeq 
  \varepsilon_K +\varepsilon'- \delta_{K}
\end{equation}
\par
where $\varepsilon_K$ is the CP violation parameter in the mixing, 
$\varepsilon'$ is the direct CP violation parameter 
and $\delta_K$ stands for the amount of CPT violations. 
In the Standard Model Extension (SME) framework 
\cite{Kostelecky:2001ff}, according to Greenberg theorem 
\cite{Greenberg:2002uu},   
the $\delta_K$ parameter is  expected to have the following expression:
\begin{equation} \label{eq:intro:deltak}
\delta_K \approx i \sin \phi_{SW} e^{i \phi_{SW}} \gamma_K (\Dao-
\vec{\beta}_K \cdot\Delta{\vec{a}})/\Delta m,
\end{equation} 
where $\gamma_K$ and $\beta_K$ are the usual Lorentz factors for the
kaon, $\phi_{SW}$ is the super-weak phase and $\Damu$ are the SME parameters 
for the kaon system. 
\par
\Equ{eq:intro:deltak} shows that $\delta_K$ is modulated by the kaon 
momentum modulus ($\gamma_K$ and $|\vec{\beta}_K|$) and by its spatial 
direction ($\vec{\beta}_K$).   
In the KLOE case the two kaons are produced almost back-to-back in the 
\f\ decay and therefore evolve with two different $\delta_K$
($\delta_K(\vec{P}_1)\not=\delta_K(\vec{P}_2)$). Additional angular 
dependence in the \equ{eq:intro:timeevo} through \equ{eq:intro:deltak} 
are induced by the Earth motion (sidereal time variation) and residual \f\ 
momentum in the lab frame.  
The effect produced by CPT violation can be 
observed in the distribution of \equ{eq:intro:timeevo} provided that the 
two kaon final states are tagged with respect to some reference direction in 
the laboratory frame and taking into account the proper coordinate 
transformation in the privileged reference frame where \equ{eq:intro:deltak} 
holds in this form (see equation 14 of reference \cite{Kostelecky:2001ff}).
The signal selection starts withe the request
of two decay vertices with two tracks connected each. 
The same kinematical selection criteria are requested in order to 
obtain the cleanest and less biased sample of $K\to\pi\pi$ possible.
At the end of the selection chain the background contamination,
according to the MC estimate, is 1.5\% and is mainly due to kaon 
regeneration on the beam pipe structure.
\par
The data distributions 
have been fitted with the \equ{eq:intro:timeevo} including the 
SME effects. 

\begin{figure}[htb]
  \begin{center}
    \includegraphics[width=0.50\textwidth]{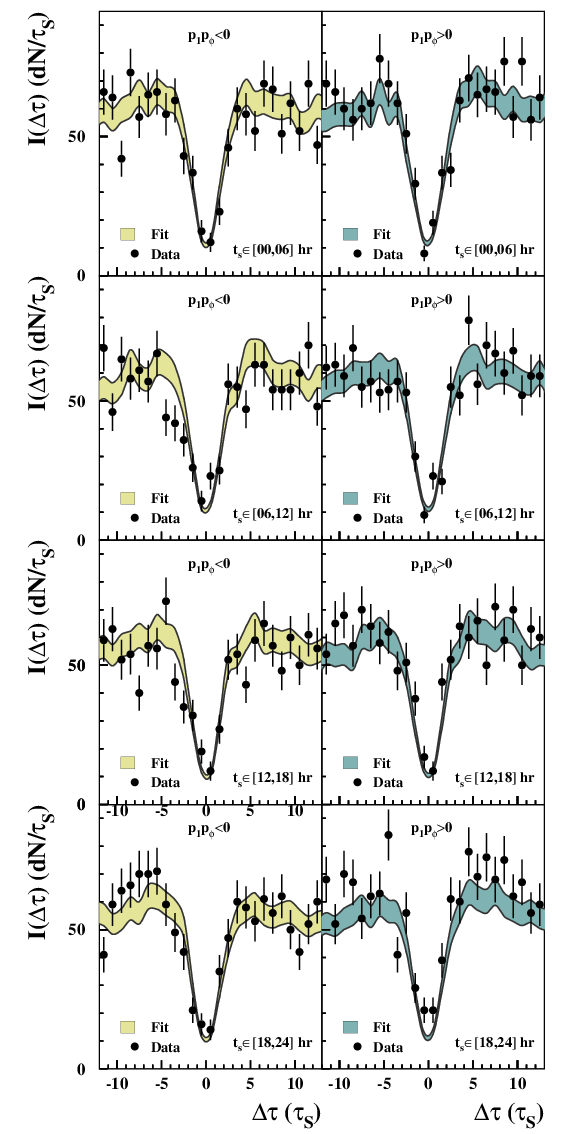}
  \end{center}
  \caption{Fit results: the top and bottom plots refer to the two 
    angular selections. Black points are for experimental data while 
    colored bands are the fit output. 
    The error bars on experimental data are purely statistical, while the band 
    represents the contribution to the uncertainty due to MC 
    simulation statistics and efficiency correction.
    The fit $\chi^2/N_\mathrm{DoF}$ is $211.7/187$ corresponding 
    to a probability of 10\%.} 
  \label{fig:results:datafit}
\end{figure}

The \Dt\ range is: $\Dt \in [-12:12]\tau_S$. This choice is to limit 
the perturbation on the result by the regeneration on the thicker 
part of the beam pipe, the Al-Be spherical pipe\cite{Ambrosino:2006vr}. 
The sample of data collected has been splitted in four sidereal time 
bins and two angular selection for a grand total of 192 experimental 
points simultaneously used for fitting, as shown in \fig{fig:results:datafit} 
\par
The results for the \Damu\ parameters are \cite{Babusci:2013gda}:\\
\begin{center}
$\Dao = (-6.0 \pm 7.7_{stat} \pm 3.1_{syst})\times 10^{-18} \,\, \GeV$,  \\
$\Dax = (\,\,\,0.9 \pm 1.5_{stat} \pm 0.6_{syst})\times 10^{-18} \,\, \GeV$,  \\
$\Day = (-2.0 \pm 1.5_{stat} \pm 0.5_{syst})\times 10^{-18} \,\,\GeV$,  \\
$\Daz = (\,\,\,3.1 \pm 1.7_{stat} \pm 0.5_{syst})\times 10^{-18} \,\,\GeV$.\
\end{center}
Which constitute the most precise result in the quark sector of the SME.
The total error is fully dominated by the 
statistical uncertainty.

\subsection{BR($K^+\to\pi^+\pi^-\pi^+(\gamma)$) measurement}
The last measurement of BR($K^{\pm}\rightarrow \pi^{\pm}\pi^+ \pi^-)$ was performed in 1972 without information 
about the radiation cut-off~\cite{chiang}
whereas value reported in PDG~\cite{PDG}
is obtained from a global fit that does not use any of the available BR($K^{\pm}\rightarrow \pi^{\pm}\pi^+ \pi^-)$ 
measurements
but the rate measurement $\Gamma(\pi^+\pi^+\pi^-) = (4.511 \pm 0.024)\times 10^6$~s$^{-1}$
published in 1970~\cite{Ford}.
In the $\pi^0\pi^0$ invariant mass distribution of data collected by NA48 a cusp-like anomaly at
$M_{00} = 2m_{\pi^+}$ can be observed~\cite{NA48-1}. The interpretation is the final state charge-exchange reaction
$\pi^+\pi^- \rightarrow \pi^0\pi^0$ in $K^{\pm} \rightarrow \pi^{\pm}\pi^+\pi^-$ decay~\cite{Budini,Cabibbo}.
Based on the fit of models~\cite{Cabibbo-Isidori} and~\cite{Berna1,Berna2} to experimental $M_{00}^2$ distribution
the difference
between the S-wave $\pi\pi$ scattering lengths in the isospin $I$=0 and $I$=2 states was determined~\cite{NA48-2},
where the main source of uncertainty
was due to the ratio of the branching ratios $K^{\pm} \raE \pi^{\pm}\pi^-\pi^+$ and $K^{\pm} \raE \pi^{\pm}\pi^0\pi^0$.

The new measurement performed at KLOE~\cite{Babusci:2014hxa} is based 
on the two samples selected with the usage of ~\Kmunu ($K_{\mu2}$ tags) and ~\Kpipi ($K_{\pi2}$ tags) events.
These independent samples of pure kaons for the signal selection 
are useful for systematic uncertainties evaluation and cross-checks~\cite{KpmSemil}.
The above mentioned decays are identified from the momentum of the charged secondary
track in the kaon rest frame evaluated using the pion mass hypothesis,
the selection efficiency of the two tagging normalization samples are similar and about 36\%.
In order to reduce influence of the trigger efficiency on the signal side, a normalization sample of $K_{\mu2}$ or $K_{\pi2}$ tags is selected.
In addition $K^-$ is used as the tagging kaon ($K_{\mu2}$ or $K_{\pi2}$) and $K^+$ as the tagged kaon (signal) due to a
factor of $\sim 10^3$ lower nuclear cross section for positive kaons with momenta $\sim$100MeV with
respect to that of negative kaons~\cite{Knuclear}.

To evaluate the momentum of the tagged kaon at the interaction point (IP) the momentum of the tagging kaon at the IP 
(backward extrapolated from its first hit in the
drift chamber (DC)) and the momentum of the $\phi$-meson measured run by run with Bhabha scattering events are used.
Then momentum of $K^+$ is extrapolated inside the DC (path of the signal kaon).
Requirement of the position of $K^+$ decay vertex inside inner radius of the drift chamber is applied in order to decrease number of charged track
in the DC, since both kaon and pions have momenta lower than $\sim200$MeV and therefore  curl up in the KLOE magnetic field, which increases
the number of tracks reconstructed with low quality. Additionally only two of the pion tracks are reconstructed to search for a vertex along the signal kaon path.

The number of $K^+\raE \pi^+\pi^-\pi^+(\gamma)$ is extracted from the
comparison of the MC-predicted shapes for the signal and the background and the experimental 
missing mass spectrum $m^2_{miss} = E^2_{miss} - (\vec{p}_{K^+} - \vec{p}_1 - \vec{p}_2)^2$
where $\vec{p}_1$ and $\vec{p}_2$ are the momenta of the selected tracks.

The branching ratio is given by:
\begin{equation}
BR(K^+ \raE \pi^+\pi^-\pi^+(\gamma)) = \frac{N_{K\raE3\pi}}{N_{tag}}\times\frac{1}{\epsilon}
\end{equation}
where $N_{K\raE3\pi}$ is the number of signal events, $N_{tag}$ is the number of tagged events and $\epsilon$ is
the overall signal selection efficiency (detector acceptance, reconstruction efficiency, tag bias, corrections for the machine and cosmic-ray background).

The effect of the influence of the charged kaon lifetime through the detector acceptance on the BR($K^+\rightarrow \pi^+\pi^+ \pi^-(\gamma)$) 
was also evaluated based on the MC simulation and then applied as a weight to the MC events, both for the signal and the control sample
selection procedures.

Averaging two results for $K^- \raE \mu^- \bar{\nu} (\gamma)$ and $K^- \raE \pi^- \pi^0 (\gamma)$ samples, accounting for correlations, the final result of 
the measurement is~\cite{Babusci:2014hxa}:
$$
BR(\kp\to\pip\pim\pip(\gamma)) = (55.65 \pm 0.31_{sta} \pm 0.25_{sys})10^{-3}
$$
which is fully inclusive of final-state radiation and has a 0.72$\%$ accuracy, 
which makes it a factor $\simeq$ 5 better with respect to the previous direct 
measurement~\cite{chiang}.
The PDG \cite{PDG} value, obtained from a constrained fit is:
$$
BR(\kp\to\pip\pim\pip(\gamma)) = (55.9 \pm 0.4)10^{-3}.
$$

\subsection{Dark forces searches}
A series of recent astrophysical observations have obtained results which cannot be explained within the
framework of the Standard Model (SM) \cite{dark1,dark2,dark3,dark4,dark5,dark6,dark7} and 
\cite{dark8,dark9,dark10,dark11}. 
Several dark matter models have been proposed to explain these anomalies. 
In particular a new gauge interaction would be mediated by a new massive vector gauge boson, the U boson
(dark photon), which could kinetically mix with the SM hypercharge (ordinary photon). 
The mass of the U boson is expected to be less than two proton masses to avoid effects, not observed, 
to the anti-proton astrophysical flux. 
This small coupling between dark matter and the SM can be described by a single kinetic mixing parameter:
$\epsilon(=\alpha_D/\alpha_{EW})$. The resulting
Lagrangian would be:
$$
\mathcal{L}_{mix} = -\frac{\epsilon}{2}F_{\mu\nu}^{EW}F_{Dark}^{\mu\nu}
$$
The U bosons should be observed as a sharp resonance at $m_U$ in the invariant-mass
distributions of final-state charged lepton or pion pairs in reactions of the type 
$\epm\to U\gamma$ with $U\to X^+X^- (X=e/\mu/\pi)$ or in meson Dalitz decays.
\par
\begin{figure}[htb]
  \begin{center}
    \includegraphics[width=0.50\textwidth]{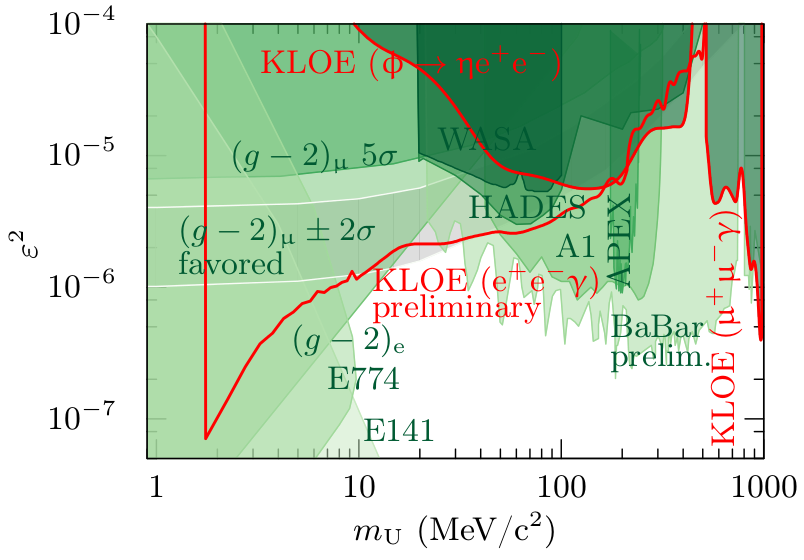}
  \end{center}
  \caption{Summary of 90\% CL exclusion limits on the dark sector 
    coupling as a function of the U boson invariant mass.} 
  \label{dark:all}
\end{figure}

In Ref.~\cite{dark12} \f\ decay like $\f\to\eta U$ with $U\to\epm$ are proposed.
At KLOE two different searches were performed using the decay chain $\f\to\eta U$ with $U\to\epm$ and
$\eta\to\pip\pim\pio$ \cite{dark13} and $\eta\to\pio\pio\pio$ \cite{dark14}. 
The two analyses selected a final sample of $\sim 13000$ 
and $\sim 31000$ events, respectively, using 1.7 \invfb of data. Irreducible background from Dalitz decay
$\f\to\eta\epm$ was simulated in the assumption of VMD models \cite{dark15}. A resonant peak was not observed
and the CLS technique was used to set an upper limit on the strength of kinetic mixing parameter
as a function of U boson mass \cite{dark16}. The 90\% confidence
level limit is shown in \fig{dark:all}.
\par
Using 239.3 \invpb\ of data collected in 2002 a search for U boson in the process 
$\epm\to U\mu^+\mu^-$ has been performed \cite{dark17}. The signal would appear as a narrow resonance in the 
final state dilepton invariant-mass spectrum.
For this analysis the two charged tracks are required to be emitted at large-angle such that their 
energy is deposited in the barrel of the calorimeter. The initial-state radiation (ISR) photon was 
not explicitly detected, being emitted at small angle with respect to the beam axis,
where its direction was reconstructed using kinematics of the charged leptons 
The signal selection is performed by using a variable called the “track mass”,
$m_{track}$, that was computed using energy and momentum conservation, assuming two 
equal-mass oppositely-charged final-state particles and one unobserved photon.
The final invariant mass spectrum was obtained after subtracting residual 
background and correcting for efficiency and luminosity.
No resonant peak was observed so the CLS technique was used to estimate the maximum 
number of U boson signal events that can be excluded at 90\% confidence level, $N_{CLs}$.
In the estimate a systematic uncertainty of $\sim 1.5\%$ has been taken into account. 
From $N_{CLs}$ is possible to estimate a limit on the kinetic mixing parameter:
\begin{equation}
\label{dark:isrlimit}
\epsilon(m_{\mu\mu}) = \frac{\alpha_D}{\alpha_{EW}}=
\frac{N_{CLs}(m_{\mu\mu})}{\varepsilon_{eff}(m_{\mu\mu})}
\frac{1}{H(m_{\mu\mu})I(m_{\mu\mu})\mathcal{L}_{int}}
\end{equation}
where the radiator function, $H(m_{\mu\mu})$, was evaluated with MC dedicated 
simulation based on PHOKARA  and $\mathcal{L}_{int}$ = 239.3\invpb\ 
is the integrated luminosity and $\varepsilon_{eff}(m_{\mu\mu})$ ranges between 
1\%-10\%.
The 90\% confidence level limit is shown in  \fig{dark:all}.
\par
The search of $U\to\mu^+\mu^-$ is obviously limited to U boson masses with 
$m_U>2m_\mu$. To scan for lower values a search in the \epm\ final state is
needed. In this case, to have enough statistics around the threshold ($m_{ee}=2m_e$),
the event selection has been performed by selecting the ISR photon and 
leptons emitted at large angle (detected in the barrel).  
The resulting background contamination
was less than 1.5\%.
No resonant U boson peak was observed prompting another use of the 
CLS technique to estimate $N_{CLs}$, the
number of U boson signal events excluded at 90\% confidence
level. We used \equ{dark:isrlimit} with $m_{\mu\mu} \to m_{ee}$ 
to set a preliminary limit on the kinetic mixing parameter as a function
of $m_U$. For this analysis $\varepsilon_{eff}$ ranges between 1.5\% and 2.5\%, and
the $\mathcal{L}_{int}=1.54~\invfb$ from 2004-05 KLOE data.
Results are shown in \fig{dark:all}
\par
\begin{figure}[htb]
   \begin{center}
     \includegraphics[width=0.40\textwidth]{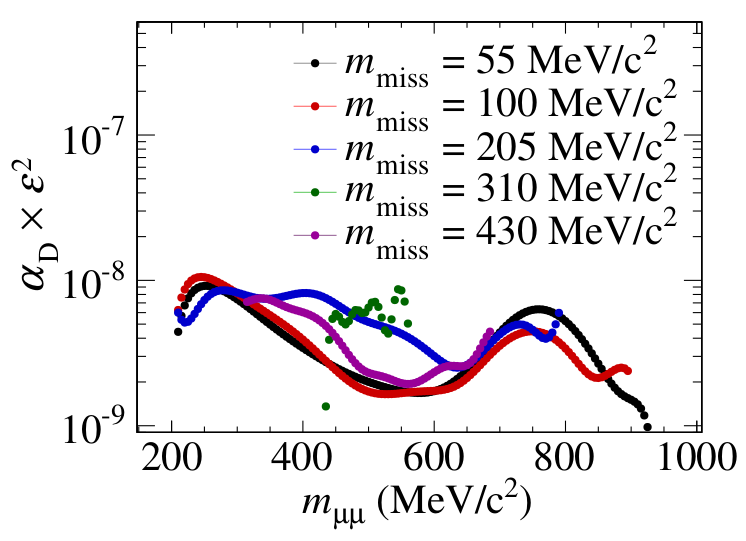}
   \end{center}
   \caption{Limits from dark Higgs-strahlung process.} 
   \label{dark:higgs}
 \end{figure}

At KLOE it has been investigated also a scenario in which the hidden symmetry 
is spontaneously broken by a Higgs-like mechanism, thus implying the 
existence of at least one other scalar particle, $h'$. 
The hypothetical dark Higgs-strahlung process $\epm\to U h'$ with 
$U\to\mu\mu$ 
has been investigated using KLOE data. The main advantage of this 
process is that is suppressed by a single factor of $\epsilon$ instead
of $\epsilon^2$ as for the process already described. The obvious implication
of that is the possibility to set stricter limits with same amount of 
statistics. 
The production cross section of this process would be proportional
to the product of the dark coupling and the kinetic mixing 
strength $\alpha_D\times\epsilon^2$ \cite{dark27}.
Assuming $m_{h'}<m_U$ the dark Higgs boson would have a large
lifetime escaping the KLOE detector without interacting. 
KLOE has performed this analysis using 1.65 \invfb\ of
data collected with center-of-mass collision energy at
the  \f-peak, and 0.2 \invfb\ of data with a center-of-mass
energy of  1000 MeV. Mass resolutions were found to
be  1 \MeV\ for $m_{\mu\mu}$ ($m_U$) and  10 \MeV\ for the missing mass 
($m_{h'}$). The signature of the dark Higgs-strahlung process
would be a sharp peak in the two-dimensional distribution
missing mass versus $m_{\mu\mu}$. The binning has been chosen
such that 90-95\% of the signal would be in a single bin.
The evaluation of background has been performed using a 
sliding $5 \times 5$ bin matrix (excluding the central bin) to determine 
background MC simulation scale factors. 
The selection efficiency was evaluated using MC simulations and varied 
between 15\% and 25\%. A conservative estimate of 10\% systematic contribution
has been considered.
No evidence of the dark Higgs-strahlung process was found. 
Using uniform prior distributions, 90\% confidence level Bayesian upper 
limits on the number of events, $N_{90\%}$, were derived separately for the two 
samples used. The results were then converted in terms of the dark Higgs-strahlung 
production cross section parameters:

$$
\alpha_D\times\epsilon^2 = \frac{N_{90\%}}{\varepsilon_{eff}}
\frac{1}{\sigma_{Uh'}(\alpha_D\epsilon^2=1).\mathcal{L}_{int}}
$$
where $\mathcal{L}_{int}$ is the integrated luminosity and
$$
\sigma_{Uh'}\propto \frac{1}{s}\frac{1}{(1-m_U^2/s)^2}
$$
is the total cross section evaluated in the assumption $\alpha_D\epsilon^2=1$.
The combined 90\% confidence level limits from on- and off-peak data are
shown \fig{dark:higgs} as projections onto $m_{\mu\mu}$.

\section{Conclusions}

\dafne\ commissioning has been delayed by several external faults, 
but several clear results have been achieved in terms of instantaneous 
luminosity and beam currents.
Limiting factors to achieve the desired goal of 3 \invfb/year integrated 
luminosity have been investigated and several knobs still have to be 
used to fully exploit \dafne\ capability.
KLOE-2 detector commissioning is on­going and the collaboration is 
quite confident to be able to start data taking during next months.
Performance of new detectors as well as the  old one are as expected.
The KLOE-2 collaboration is still exploiting the old KLOE dataset producing 
several interesting physics results.
The new dataset will be of great importance to improve and complete 
several KLOE measurements especially on the $\gamma\gamma$ physics 
and kaon interferometry.

\end{document}